\documentstyle [12pt,aaspp4]{article}
\def\beq{\begin{equation}}
\def\eeq{\end{equation}}
\def\bey{\begin{eqnarray}}
\def\eey{\end{eqnarray}}
\def\pppm{\rm P^3M}
\def\mpc{\,h^{-1}{\rm {Mpc}}}

\def\gs{\mathrel{\raise1.16pt\hbox{$>$}\kern-7.0pt
\lower3.06pt\hbox{{$\scriptstyle \sim$}}}}
\def\ls{\mathrel{\raise1.16pt\hbox{$<$}\kern-7.0pt
\lower3.06pt\hbox{{$\scriptstyle \sim$}}}}
\def\gtsima{$\; \buildrel > \over \sim \;$}
\def\ltsima{$\; \buildrel < \over \sim \;$}
\def\prosima{$\; \buildrel \propto \over \sim \;$}
\def\gsim{\lower.5ex\hbox{\gtsima}}
\def\lsim{\lower.5ex\hbox{\ltsima}}
\def\simgt{\lower.5ex\hbox{\gtsima}}
\def\simlt{\lower.5ex\hbox{\ltsima}}
\def\simpr{\lower.5ex\hbox{\prosima}}

\begin{document}
\title {
The density profile of equilibrium and
non-equilibrium dark matter halos}
\author {Y.P. Jing} 
\affil{Research Center for the Early Universe,
School of Science, University of Tokyo, Bunkyo-ku, Tokyo 113-0033, Japan}
\affil {e-mail: jing@utaphp2.phys.s.u-tokyo.ac.jp}
\received{---------------}
\accepted{---------------}

\begin{abstract}
  We study the diversity of the density profiles of dark matter halos
  based on a large set of high-resolution cosmological simulations of
  $256^3$ particles. The cosmological models include four scale-free
  models and three representative cold dark matter models. The
  simulations have good force resolution. In each cosmological model,
  there are about 400 massive halos which have more than $10^4$ particles
  within the virial radius. Our unbiased selection of all massive
  halos enables us to quantify how well the bulk of dark matter halos
  can be described by the Navarro, Frenk \& White (NFW) profile which
  was established for equilibrium halos. We find that about seventy
  percent of the halos can be fitted by the NFW profile with a fitting
  residual ${\rm dvi_{max}}$ less than 0.3 in $\Omega_0=1$
  universes. This percentage is higher in the low-density cosmological
  models of $\Omega_0=0.3$.  The rest of the halos exhibits larger
  deviation from the NFW profile for more significant internal
  substructures. There is a considerable amount of variation in the
  density profile even among the halos which can be fitted by the NFW
  profile (i.e. ${\rm dvi_{max}}<0.30$). The distribution of the
  profile parameter, the concentration $c$, can be described well by a
  lognormal function with the mean value $\bar c$ slightly smaller
  (15\%) than the NFW result and the dispersion $\sigma_c$ in $\ln c$
  about 0.25. More virialized halos with ${\rm dvi_{max}}<0.15$ have
  the mean value $\bar c$ in better agreement with the NFW result and
  their dispersion $\sigma_c$ is also slightly smaller (about 0.2).
  Our results can alleviate some of the conflicts found recently
  between the theoretical NFW profile and observational
  results. Implications for theoretical and observational studies of
  galaxy formation are discussed.
\end{abstract}


\section {Introduction} 
In their recent series of papers, Navarro, Frenk, \& White (NFW;
\cite{nfw95}, \cite{nfw96}, \cite{nfw97}) investigated the density
profile $\rho(r)$ for Dark Matter (DM) halos in cosmological models of
hierarchical clustering. They concluded that the density profile has a
universal form,
\beq\label{nfwprof}
\rho_{NFW} (r)\propto {1\over {r(r+r_s)^2}}={1\over r^3_{200}{x(x+1/c)^2}}\,,
\eeq
where $r_s$ is the core radius, $r_{200}$ is the virial radius within
which the mean density is 200 times the critical density $\rho_{\rm
  crit}$, $x$ is the radius scaled with $r_{200}$, and $c\equiv
r_{200}/r_s$ is the concentration parameter. For a given virial mass
$M_{200}$, its virial radius $r_{200}$ is fixed [given by $(3
M_{200}/800 \pi\rho_{\rm crit})^{1/3}$] and the profile has only one
free parameter $c$. Their simulation results show that $c$ depends
only on the halo mass for a given cosmological model. Furthermore they
proposed a recipe based on the Press-Schechter theories with which one
can predict $c$ very accurately for any hierarchical model. All these
results have significant impact on theoretical and observational
studies of galaxy formation. The density profile has been widely used
for modeling galaxy formation (e.g. \cite{dalcantonetal97},
\cite{mmw98}, \cite{mm98},\cite{bosch98}) and for interpreting and
confronting various observational data (e.g., \cite{pss96}, Carlberg
et al. \cite{carlbergetal97a},\cite{carlbergetal97b}, \cite{mss98},
Tyson et al. \cite{tysonetal98}, \cite{mb98}, \cite{n98}).

While it is truly important to emphasize the universality of the
density profile as NFW already did, our present work will focus on
another important aspect of the density profile: diversity. The halos
NFW studied are intentionally selected to be at the equilibrium state.
NFW picked up the dark matter halos which {\it look in equilibrium}
from (lower-resolution) cosmological simulations and resimulated them
at high resolution using fine and coarse particles. They checked the
ratio of the kinetic energy to the potential energy for each halo from
redshift $z=0.1$ to the final epoch ($z=0$) and measured the density
profile at the epoch ($0.1\ge z\ge 0$) when the halo is closest to
equilibrium according to the virial theorem. With this procedure, NFW
attempted to avoid non-equilibrium DM halos. This selection procedure
however left two important questions to be answered. First, how much
fraction of the DM halos can be described by the NFW profile?  This
question is closely related to the well-known fact that quite a
fraction of DM halos has a significant amount of substructures;
strictly speaking, very few DM halos are in equilibrium (e.g. Jing et
al. {\cite{jingetal95}}; Thomas et al. {\cite{thomasetal98}}). Second,
even for the halos which can be fitted by the NFW profile, what is the
distribution of the profile parameter $c$?  Answering both questions
would be of vital importance for the theoretical modeling and the
observational interpretation based on the NFW profile.

To answer these questions, one needs a good sample of DM halos. The
sample must be large and well-defined, and the individual halos must
be well-resolved.  In this paper, we use a large set of
high-resolution cosmological simulations of $256^3$ particles (Jing \&
Suto \cite{js98}; Jing \cite{jing98a}) to study these problems. The
simulations were generated with our Particle-Particle/Particle-Mesh
($\pppm$) code which has high force-resolution. Four scale-free models
with $P(k)\propto k^n$ ($n=-0.5$, $-1$, $-1.5$, and $-2.0$) and three
representative CDM models are simulated. In each model, there is a
total of a few hundred massive halos with more than $10^4$ particles
within the virial radius $r_{200}$. Both the force and the mass
resolutions of our halos are comparable to those of NFW. But our
unbiased selection of all halos and the large number of the halos
enable us to answer the questions NFW left.

Our results will show that the goodness of the NFW profile fitting depends
on if the halo is in equilibrium. About $15$ to $40$ percent of DM
halos, depending on the density parameter $\Omega_0$, cannot be fitted
well by the NFW profile because of a significant amount of
substructures.  The rest of halos which are more close to equilibrium can
be described quite well by the NFW profile, with the fitting residual
less than 30 percent. For these halos which can be fitted by the NFW
profiles, there is a considerable dispersion in the density profiles
even for the same mass $M_{200}$ in the same cosmological model. The
dispersion is characterized by the distribution of the concentration
parameter $c$. We will, for the first time, show that the distribution
of $c$ can be well described by a lognormal function, with the mean
value of $c$ very close to the NFW result and the scatter $\sigma_c $ in
$\ln c$ about 0.25. Our result therefore essentially confirms the
conclusion of NFW, but significantly extends the study of the density
profiles to a wider range of halos with different physical properties.
We will discuss the implications of our result for theoretical and
observational studies in Sect.~4. A summary of these results already
appeared in the conference proceedings {\sl Evolution of Large-scale
  Structure} at Garching, 1998 (Jing \cite{jing98b}).

In the next section (\S 2) we will discuss our simulations. Our
results for the density profiles will be presented in \S3. We will
summarize our results and discuss their implications for the studies
of galaxy formation in \S 4.

\section{Simulations}

We use a set of high-resolution simulations of $256^3$ particles which
were generated with our vectorized $\pppm$ code on the supercomputer
Fujitsu VPP300/16R at the National Astronomical Observatory of
Japan. The simulations were evolved typically by 1000 steps, with a
force resolution $\epsilon$ ($\epsilon$ is the force softening of the
Plummer form) about $2\times 10^{-4}$ of the simulation box size. We will
use only the final output of each simulation for the present study.
The simulations cover three representative Cold Dark Matter (CDM)
models and four Scale-Free (SF) models of hierarchical clustering. The
CDM models are specified with the density parameter $\Omega_0$, the
cosmological constant $\lambda_0$, the shape $\Gamma$ and the
normalization $\sigma_8$ of the linear power spectrum $P(k)$. The SF models
assume an Einstein--de Sitter universe (i.e.  $\Omega_0=1$ and
$\lambda_0=0$) and a power-law $P(k)\propto k^n$ for the linear
density power spectrum. The amplitude of $P(k)$ for the SF models are
set by the non-linear mass $M_\ast$ at which the rms linear density
fluctuation $\sigma(M_\ast)$ is 1.68. Table~1 summarizes the model
parameters.  The simulations have been used to study the theoretical
significance of the strong clustering of high-redshift galaxies
(Steidel et al. \cite{steideletal98}) by Jing \& Suto (\cite{js98})
and to derive the accurate fitting formula for the halo clustering by
the author (Jing \cite{jing98a}; \cite{jing99}). We refer readers to
these papers for complementary information about these simulations.

\section{The distribution of density profiles}

We select DM halos using the Friends-of-Friends method with the
linking length $0.2$ times of the mean particle separation. We compute
the gravitational potential for each halo particle, and choose the
particle with the minimum potential as the center of the halo. We
calculate the density $\rho(r)$ in shells with equal logarithmic thickness
$\log_{10} \Delta r =0.1$ from $r_{200}$ inward to about $\epsilon$, a
formal force softening limit. Only the halos more massive than $10^4$
particles within the virial radius $r_{200}$ are studied here.  The
density profiles are fitted with Eq.~(\ref{nfwprof}) to get the
parameter $c$ with an equal logarithmic weighting, i.e.
\beq\label{fitmin}
\min \sum_{i} [\ln \rho(i)-\ln \rho_{NFW}(r_i)]^2 \,,
\eeq
where $\rho(i)$ is the simulation value at the $i$-th radial bin, and
$\rho_{NFW}(r_i)$ is given by Eq.~({\ref{nfwprof}) at $r_i$. We have
  tested the robustness of our fitting result with a Poisson error
  weighting and found essentially the same result. Since the Poisson
  error is negligible compared to the deviation of  the fitted
  profile from the simulation profile, we prefer to use the equal
  logarithmic weighting. In the fitting, we conservatively set the
  lower radial limit to $3\epsilon$. The goodness of the fitting will
  be characterized by the maximum relative deviation of the simulation
  $\rho(r_i)$ from the fitted $\rho_{NFW}(r_i)$ in all radial bins
  $\{i\}$, i.e.
\beq
{\rm dvi_{max}}=\max
\{|(\rho(r_i)-\rho_{NFW}(r_i))/ \rho_{NFW}(r_i)|\}\,.
\eeq

We illustrate our measured density profiles in Figure~\ref{dens-prof}
for two CDM models and two SF models. (We do not plot our results for
{\it all} models, because what we have shown is typical and the other
models just give qualitatively the same result. This way however saves
the space and we will follow this convention throughout the paper. Our
results for all models are summarized in Table~2). The profiles are
selected randomly under the condition that their maximum deviations
${\rm dvi_{max}}$ must be around $0.1$, $0.2$ and $0.6$ respectively
in each model. It is important to recall that our fitting is made for
$3\epsilon <r<r_{200}$; we plot results for $r<3\epsilon$ for the sake
of discussion in \S 4. It is interesting to note that the density
profiles for $ {\rm dvi_{max}}\approx 0.1$ and $0.2$ can all be fitted
quite satisfactorily with the NFW profile. For the cases of $ {\rm
  dvi_{max}}\approx 0.6$, the NFW profile does not seem to provide a
nice description to the simulation data. With a closer look at the figure, 
we can easily find that there is one feature common to all profiles
with $ {\rm dvi_{max}}\approx 0.6$: relative to the NFW fit, the halo
density is higher than the fitted curve at $r/r_{200}\ls 0.03$ (though
the simulation result may be underestimated because of the softening
at this radius) but lower at $r/r_{200}\ls 0.3$. The shoulder at
$r/r_{200}\approx 0.3$ of these profiles can be easily explained as
these halos having significant substructures. We have checked the
particle distributions of these halos at earlier epoches, and indeed
found that they always suffered a violent merging very recently.

With substructure measures, we can quantitatively discuss the relation
between the fitting quality and the substructures.  Here we use two
indicators for the formation history which determines substructures
(e.g. \cite{evrard93}). One is the ratio $M_{05}/M$, where $M$ is the
halo mass and $M_{05}$ is the mass of its largest progenitor at
redshift $z=0.5$. The other is the redshift $z_{05}$ at which the
halo's largest progenitor has reached half of its final mass. The
amount of substructures, on average, is a decreasing function of
either of these two quantities. In Figure~{\ref{merge}}, we plot the
maximum deviation ${\rm dvi_{max}}$ versus $M_{05}/M$ or $z_{05}$ for
halos in the SCDM and OCDM models.  The strong correlations between
the ${\rm dvi_{max}}$ and $M_{05}/M$ or $z_{05}$ confirm that the
fitting quality of the NFW profile depends on the formation history,
or equivalently the amount of the substructures.  The large
dispersions seen in the Figure~{\ref{merge}} reflect the fact that the
both formation history indicators are instant measures at some epoch,
but the substructures actually depend on the whole merging history. A
better substructure indicator would likely give a much tighter
correlation, but quantitatively confirming this statement is beyond
the scope of this paper.

We present in Figure~{\ref{cscatter}} the fitted values of $c$ for
halos in four models, two SF models and two CDM models.  Different
symbols are used to indicate the quality of the NFW fit. The circles
are for halos with ${\rm dvi_{max}}<15\%$, the triangles for $15\%
<{\rm dvi_{max}}<30\%$, and the crosses for ${\rm
  dvi_{max}}>30\%$. With this classification, it is fair to say that
the first subset of halos can be fitted by the NFW profile very well,
the second reasonably well, and the third is ill
described by the NFW profile because of significant substructures. In
order to compare with the NFW results, we plot the NFW prediction
$c_{NFW}$ based on their recipe given by Navarro et al.  (1997) (the
solid lines). The figure assures that the equilibrium halos with ${\rm
  dvi_{max}}<15\%$ are in good agreement with the NFW prediction, but
the less virialized halos have systematically smaller values of $c$.

The Probability Distribution Function (PDF) of the fitted $c$ is
presented in Figure~{\ref{cpdf}} for the four models. We choose
$c/c_{NFW}$ as the abscissa in order to correct for the mass
dependence of the parameter $c$, though this correction for the mass
range covered here is actually tiny. The PDF is shown separately for
the halos with different amounts of substructure. For each subset of
halos, the PDF is fitted by a lognormal function
\beq\label{lognormal}
p(c)dc={1\over \sqrt{2\pi}\sigma_c}
\exp{-{(\ln c-\ln \bar c)^2\over 2\sigma_c^2}} d\ln c\,.
\eeq
The fitted values of $\bar c$ and $\sigma_c$ are listed in Table~2 for
{\it all} the models. The fitted curves are shown in
Figure~{\ref{cpdf}}, which indicate that the lognormal function provides
an excellent fit to our simulation results of $c$.

From Table~2 we see that the most virialized halos (${\rm
  dvi_{max}}<15\%$), which are best fitted by the NFW profile, have a
mean concentration $\bar c$ which also agrees very well with the NFW
prediction. The difference between our $\bar c$ and the NFW prediction
is less than 20\% in all models. A similar amount of difference also
existed in the original work of NFW between their recipe for $c$
(which is used here) and their simulation result. This agreement
between our result and the NFW is not surprising, since both studies
are analyzing the equilibrium halos. But considering the vast
difference in the simulation methods between the two studies, the
agreement is very encouraging and our result gives further supporting
evidence to the NFW profile.  The less virialized halos (with larger
${\rm dvi_{max}}$) have a smaller concentration, i.e. flatter density
profile at the center. The mean concentration of the halos with
$15\%<{\rm dvi_{max}}<30\%$ is about 15 percent smaller than that of
${\rm dvi_{max}}<15\%$ (except for $n=-0.5$).

The dispersion $\sigma_c$ of $\ln c$ is about 0.2 to 0.35
for the halos with ${\rm dvi_{max}}<30\%$. The subset of the most virialized
halos (${\rm dvi_{max}}<15\%$) shows less dispersion with $\sigma_c$
between 0.17 and 0.27. There appears no clear correlation of the
fitting parameters $\bar c$ and $\sigma_c$ with the cosmological parameters
or the density power spectrum. 

The halos with significant substructures (${\rm dvi_{max}}>30\%$) have
much smaller concentration and much larger variation in density
profile than the more virialized halos. The poor fit of the NFW
profile to these halos, however, means that it is not much meaningful
to discuss the fitting parameters for this subset. Table~2 indicates
that about 35 percent of {\it all} halos is in this category in the
$\Omega_0=1$ models. Because the substructure effect is weaker in the
$\Omega_0=0.3$ models, only about 15 to 20 percent halos have ${\rm
  dvi_{max}}>30\%$.

For any simulation study, there exists a concern about the resolution
effect, i.e. how much the result is affected by limited
resolution. For the present study, such a concern is even enhanced by
the recent claims (Fukushige \& Makino 1997; Moore et al. 1998) that
$\sim 10^6$ particles are necessary to correctly sample the density
profile of a single halo. It is usually a tough task to quantify the
resolution effect, but fortunately, our newly available
high-resolution halos can serve this purpose. In an ongoing
project, we have developed a modified $\pppm$ algorithm which adopts
nested grids to simulate single halos (For clarity, we will call this
simulation as `halo simulation', and the simulations we described in
\S 2 as `cosmological simulation'). We are using this code to simulate
fifteen halos with $ \sim 2\times 10^6$ particles ($\sim 10^6$
particles end up within the virial radius). Each halo is evolved by
10000 time steps with a force resolution much higher than that of the
present paper. A Detailed discussion of the project will be published
elsewhere (Jing \& Suto, 2000). Of the fifteen halos, five
are randomly selected from the LCDM halos of the present paper.  As of
this writing, we have finished simulating two of them. One halo
has significant substructures with ${\rm dvi_{max}}=0.63$, and the
other is much more virialized with ${\rm dvi_{max}}=0.21$. The density
profiles of these two high-resolution halos are plotted in
Figure~\ref{comp}. For radius larger the resolution limit
($3\epsilon$) defined in this paper, the new density profiles are in
excellent agreement with those used in the present paper. It is more
gratifying to see that even fine substructure features of the density
profiles match very well in the both simulations, with ${\rm
  dvi_{max}}=0.69$ and $0.18$ respectively for the high-resolution
halos. We can therefore safely conclude that our measured quantities
are robust and are little affected by the simulation resolution. It is
worth pointing out that this conclusion is not inconsistent with the
studies of Fukushige \& Makino (1997) and Moore et al. (1998), since
they declare that $\sim 10^6$ particles are needed for studying the
core region with $r/r_{200}\ls 0.01$ while the present work is
studying an outer region with $r/r_{200}\gs 0.03$.  In fact, our
result of the virialized halo in Fig.~\ref{comp} supports their claims
for the core region, and the resolution tests conducted by Moore et
al. (1998) indicated as well that $\sim 10^4$ particles are sufficient
for studying the density profile at $r/r_{200}\gs 0.03$. Since the
survival time of substructures is much shorter in the core region than
in outer region, the scales relevant to the present study are
$r/r_{200}\sim 0.1$ (see Fig.~\ref{dens-prof}), therefore the
simulations used here are adequate for the present work.
Fig.~\ref{comp} also implies that the resolution limit we defined is
slightly too conservative.

\section {Conclusion and further discussion}
In this paper we have measured the density profiles for massive halos
in a large set of high-resolution cosmological simulations. Four
scale-free models and three representative CDM models are studied. We
selected all halos with more than $10^4$ particles within the virial radius
and fitted their density profiles with the NFW profile. We found,
\begin{itemize}

\item The quality of the NFW profile fitting depends
  on if the halo is in equilibrium. Substructures degrade the
  fitting quality, as expected, because the NFW profile was found for
  equilibrium halos.

\item We use the maximum deviation ${\rm dvi_{max}}$ of the simulation
  density profile from the fitted NFW profile as an indicator for the
  fitting quality. In the Einstein de Sitter universes, about 15
  percent of halos have ${\rm dvi_{max}}<0.15$ and 50 percent have
  $0.15<{\rm dvi_{max}}<0.30$. In the low density models, these
  percentages are higher, because the substructure effect is
  weaker. These numbers show that the NFW profile provides a good
  description for most of DM halos ($\gs 70\%$).

\item For the halos which can be fitted by the NFW profile, there is a
  considerable amount of dispersion in the parameter $c$. The
  lognormal function Eq.~(\ref{lognormal}) provides a very good fit to
  the distribution of $c$. Our fitted values for $\bar c$ and
  $\sigma_c$ are listed in Table~2. The mean value $\bar c$ of the
  most virialized halos ($ {\rm dvi_{max}}<0.15$) is in good agreement
  with the NFW prediction. The less virialized halos ($0.15< {\rm
    dvi_{max}}<0.30$, which can still be reasonably fitted by the NFW
  profile) have a smaller concentration with $\bar c$ about 15 percent
  smaller than that of the most virialized ones. The dispersion
  $\sigma_c$ is around 0.25, with the virialized halos being slightly
  less dispersive.
\end{itemize}

In a word, our results are in good agreement with the results of NFW
for the most virialized halos, but we extend the discussion of the
density profiles to less virialized halos which may account for most
of the halos in number. The new result of this work is that we have
quantified how much halos can really be fitted by the NFW profile and
that we have derived the important quantity, the probability
distribution function of the concentration parameter $c$.

The NFW profile has been used to analytically model the formation of
disk galaxies in the framework of hierarchical clustering
(\cite{dalcantonetal97}, \cite{mmw98}, \cite{mm98},\cite{bosch98}). In
such a approach, the Press-Schechter formalism is usually used to
calculate the abundance of halos, which means that almost all halos,
in equilibrium or with substructures, have been considered. Since our
results indicate that most of the halos can be reasonably fitted by
the NFW profile, these approaches are valid. But for the bulk of
halos, a slightly smaller concentration $c$ is preferred over
the value given by NFW that is for equilibrium halos. The dispersion of
$c$ should also be properly taken into account to survey various
properties of disk galaxies.

Our results would also be important for properly interpreting many
cosmological observations which are closely related to the density
profile of halos. The extragalactic objects (say, clusters of
galaxies, galaxies) that are observed are not guaranteed to be at the
equilibrium state. Our results, i.e. a smaller concentration $c$ for
less virialized halos and the large dispersion of $c$, can alleviate
the conflicts recently found between the observed rotation curves of galaxies,
the observed core radius of galaxies and clusters, and the predictions
based on the NFW profile (\cite{mss98}, \cite{mb98}, \cite{n98}).

Kravtsov et al. ({\cite{ketal98}}) discussed the variation of
density profiles in a way very different from our present work. They
compared the rotation curves of galactic CDM (or $\nu$CDM) halos with
a sample of observed rotation curves, and claimed a good
agreement. Comparing with the NFW density profile, however, they found
that their simulated density profiles, while having a large dispersion, are
significantly flatter (density about 5 to 10 times lower) at $r\approx 0.01
r_{200}$.  Our result does not seem to agree with their
result. Although we conservatively set the force softening to $0.03
r_{200}$, the fitted NFW profile follows very well our simulation
result down to $r\approx 0.01 r_{200}$ (see Fig.~1). We could only
expect that the force softening, if any, would have flattened our
simulated profile at $r=0.01r_{200}$. Of course we note that our halos
in the CDM models are at cluster scale, but those of Kravtsov et
al. are at galactic scale. It appears unlikely that the mass
difference can explain the difference in the density profiles, because
none of our SF models ($n=-0.5$ to $-2$) shows significantly
flatter density profile than the NFW result at the core region (also
see the results of NFW for galactic halos). We think that the
discrepancy can only be explained by the differences in the simulation
methods and/or in the ways to define the resolution limit. It is
interesting to note that the previous studies with the tree codes
(NFW, Moore et al. 1998) and with GRAPE (Fukushige \& Makino 1997) and
the present work with the $\pppm$ method have produced very consistent
results at the central core if an account is taken for the resolution
limits.

It would be also important to point out that we could not address the
important problem that density profile might be significantly steeper
than $r^{-1}$ in the very inner region of the halos (Fukushige \&
Makino 1997; Moore et al. 1998) because the resolution of our current
simulations is still limited. We are running a project to simulate
many halos with higher mass resolution ($\sim 10^6$ particles within
$r_{200}$) and force resolution ($\epsilon<0.002 r_{200}$). We shall
report these results in a future paper (Jing \& Suto, 2000).

\acknowledgments

I would like to thank Gus Evrard, Yasushi Suto and Simon White for
helpful comments and discussion, and the JSPS foundation for a
postdoctoral fellowship.  The simulations were carried out on VPP/16R
and VX/4R at the Astronomical Data Analysis Center of the National
Astronomical Observatory, Japan.

\clearpage
\begin{figure} 
\epsscale{1.0} \plotone{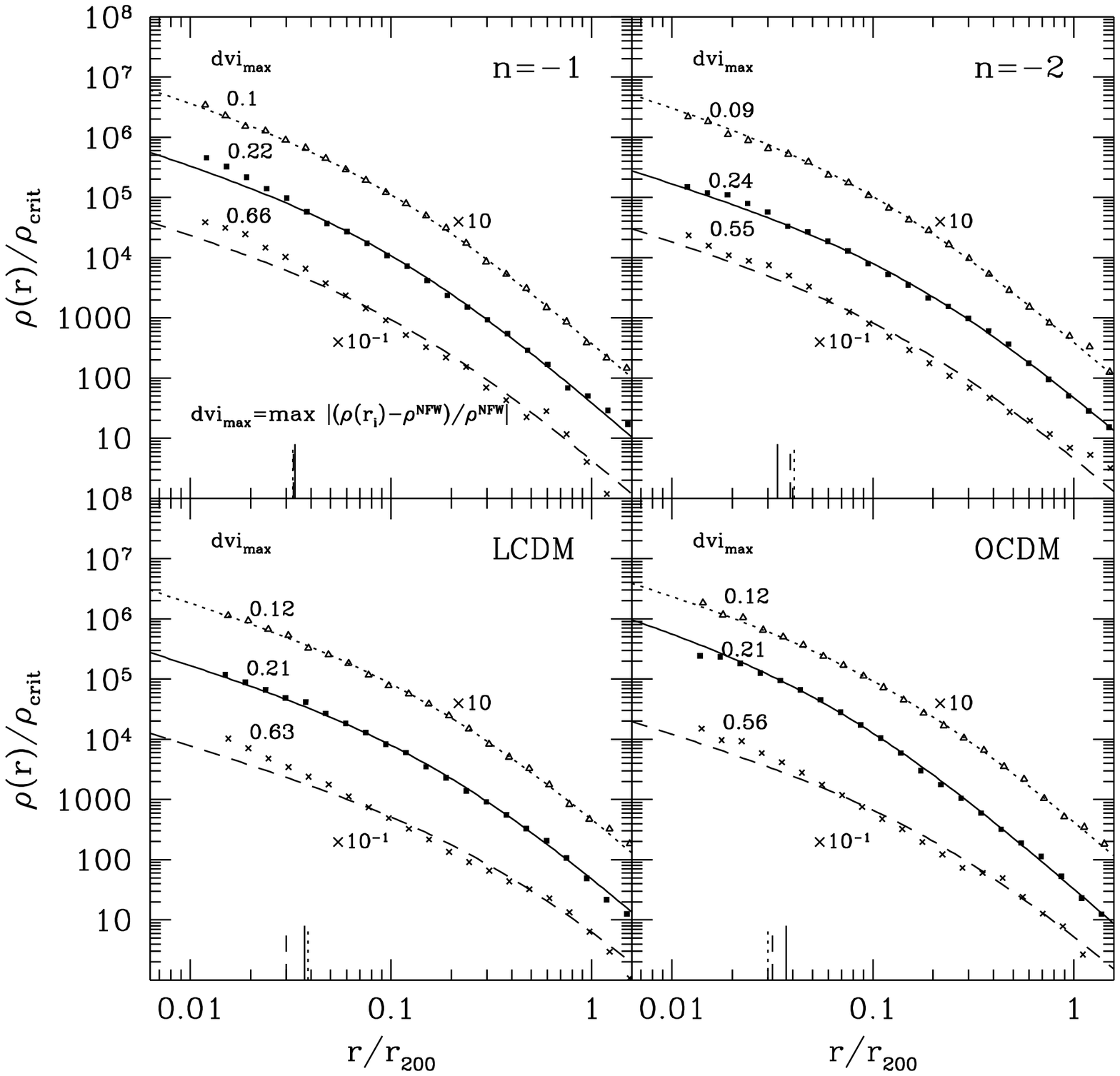} 
\caption{Illustration of halo density profiles in different
  cosmological models. Three halos are randomly selected in each model
  with the maximum deviation ${\rm dvi_{max}}$ about 0.1, 0.2, 0.6
  respectively. The exact value of ${\rm dvi_{max}}$ is labeled with a
  number below ${\rm dvi_{max}}$.  The profiles with the smallest and the
  largest ${\rm dvi_{max}}$ are shifted by one magnitude vertically, as also
  indicated in the figure. The long ticks at the bottom of each panel
  mark the lower radius limits used for the NFW fitting.}
\label{dens-prof}\end{figure}

\begin{figure}
\epsscale{1.0} \plotone{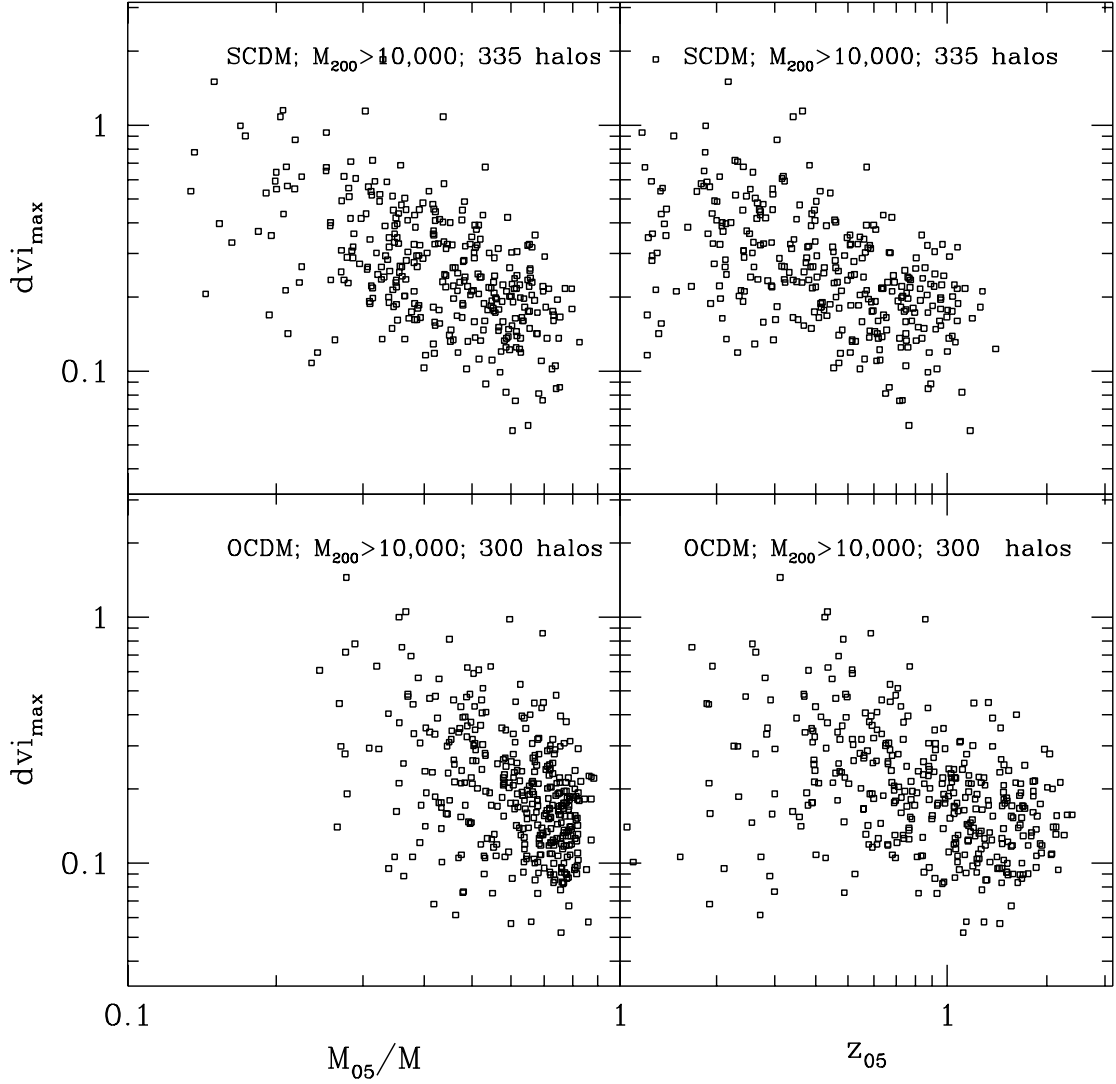}
\caption{The maximum deviation versus the formation history 
indicators $M_{05}/M$ or $z_{05}$.  }
\label{merge}\end{figure}

\begin{figure}
\epsscale{1.0} \plotone{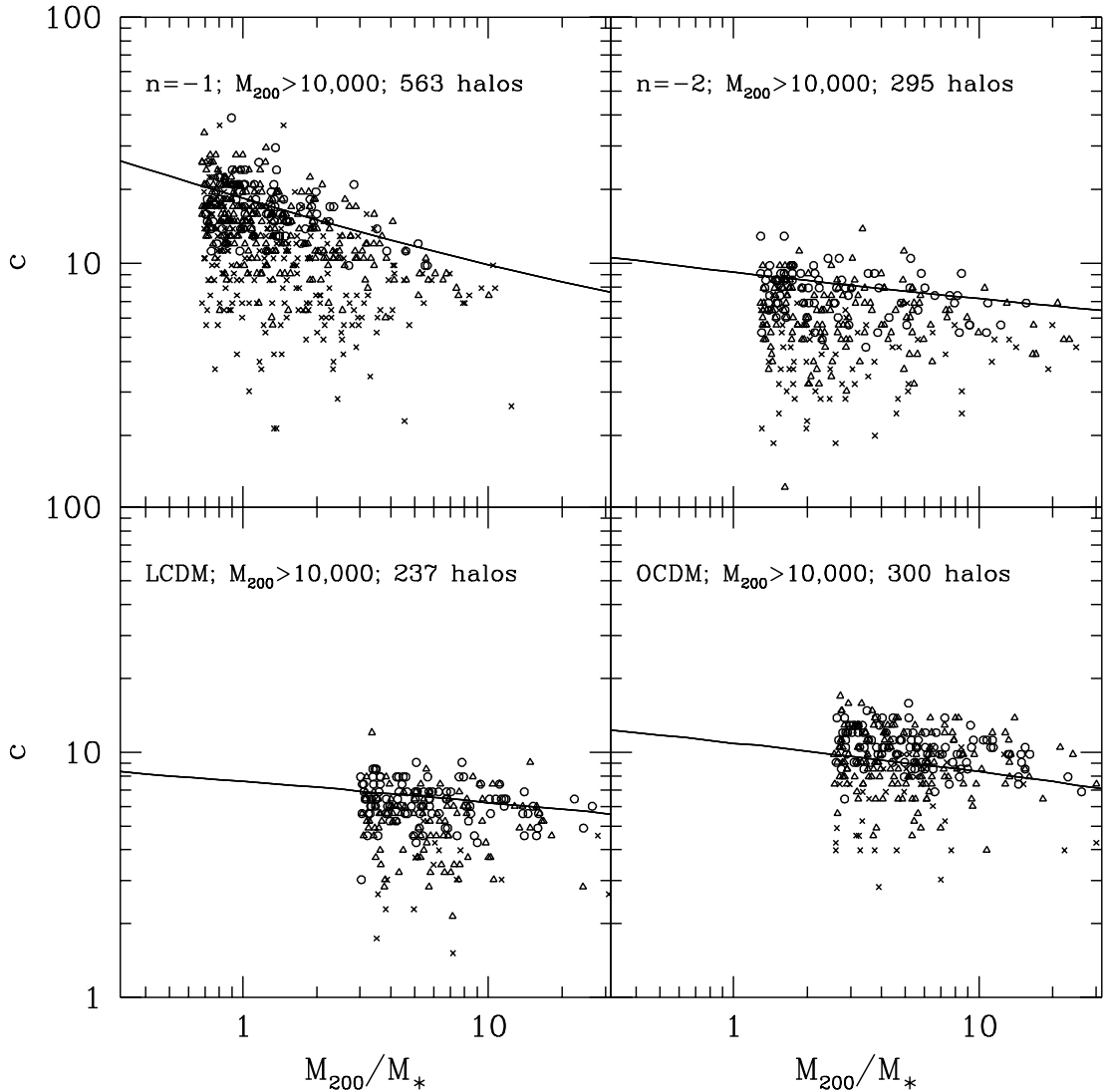}
\caption{The concentration factor $c$ is shown for four cosmological
  models as a function of the halo mass $M_{200}$. Different symbols
  are used to denote the quality of the NFW profile fit, with the
  circles for ${\rm dvi_{max}}<0.15$, the triangles for
  $0.15<{\rm dvi_{max}}<0.30$, and the crosses for ${\rm dvi_{max}}>0.30$. The
  lines are the prediction based on the NFW's recipe.
}\label{cscatter}\end{figure}

\begin{figure}{Fig.4a}
\epsscale{1.0} \plotone{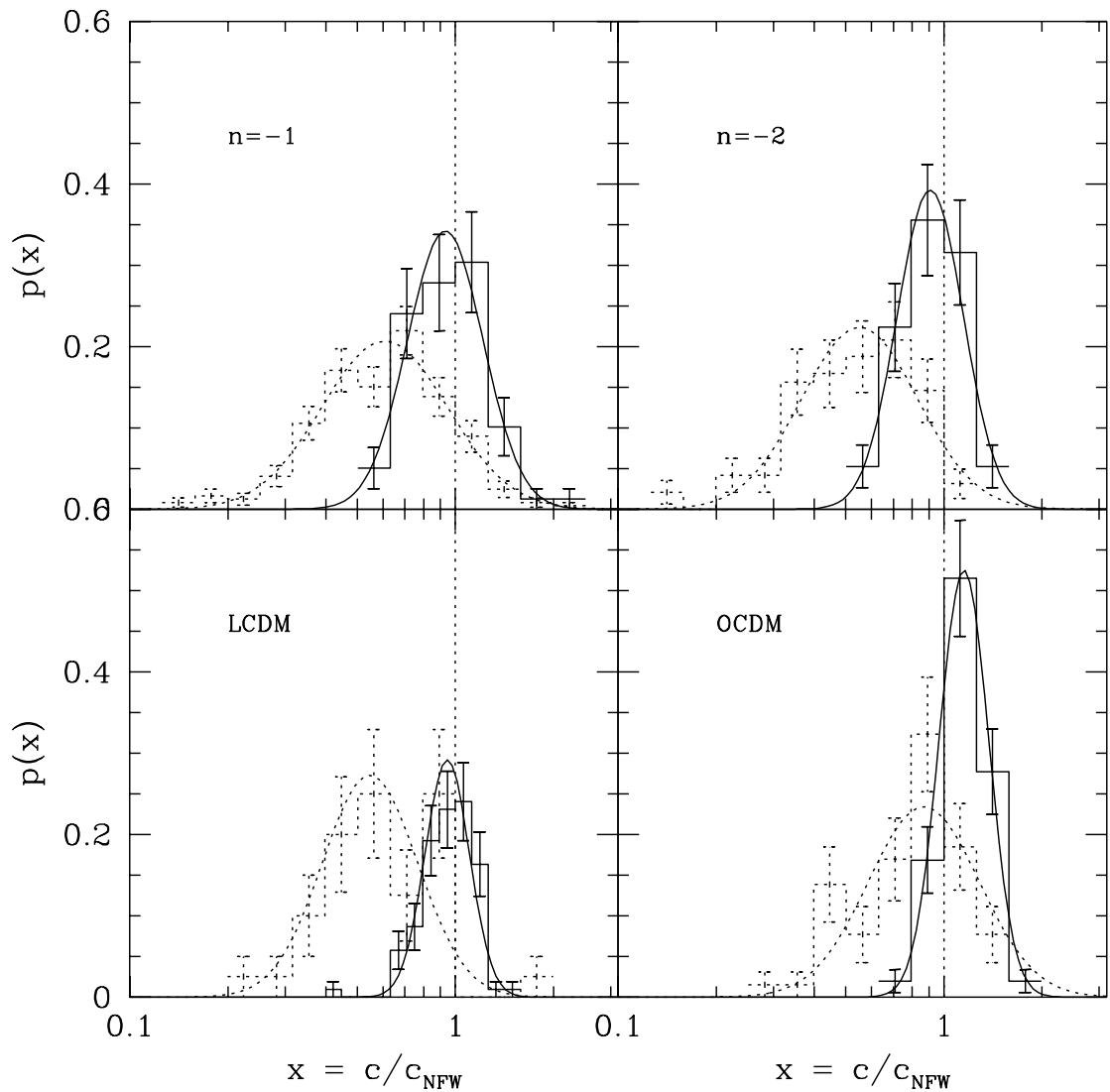}
\caption{
  The probability distribution function of the concentration parameter
  $c$ (histograms), which can be nicely fitted by lognormal
  distributions (smooth lines). {\bf a).} The solid lines are for
  the halos with ${\rm dvi_{max}}<0.15$ and the dotted ones for
  ${\rm dvi_{max}}>0.30$; {\bf b).} for $0.15<{\rm dvi_{max}}<0.30$.
}
\label{cpdf}\end{figure}
\begin{figure}{Fig.4b}
\epsscale{1.0} \plotone{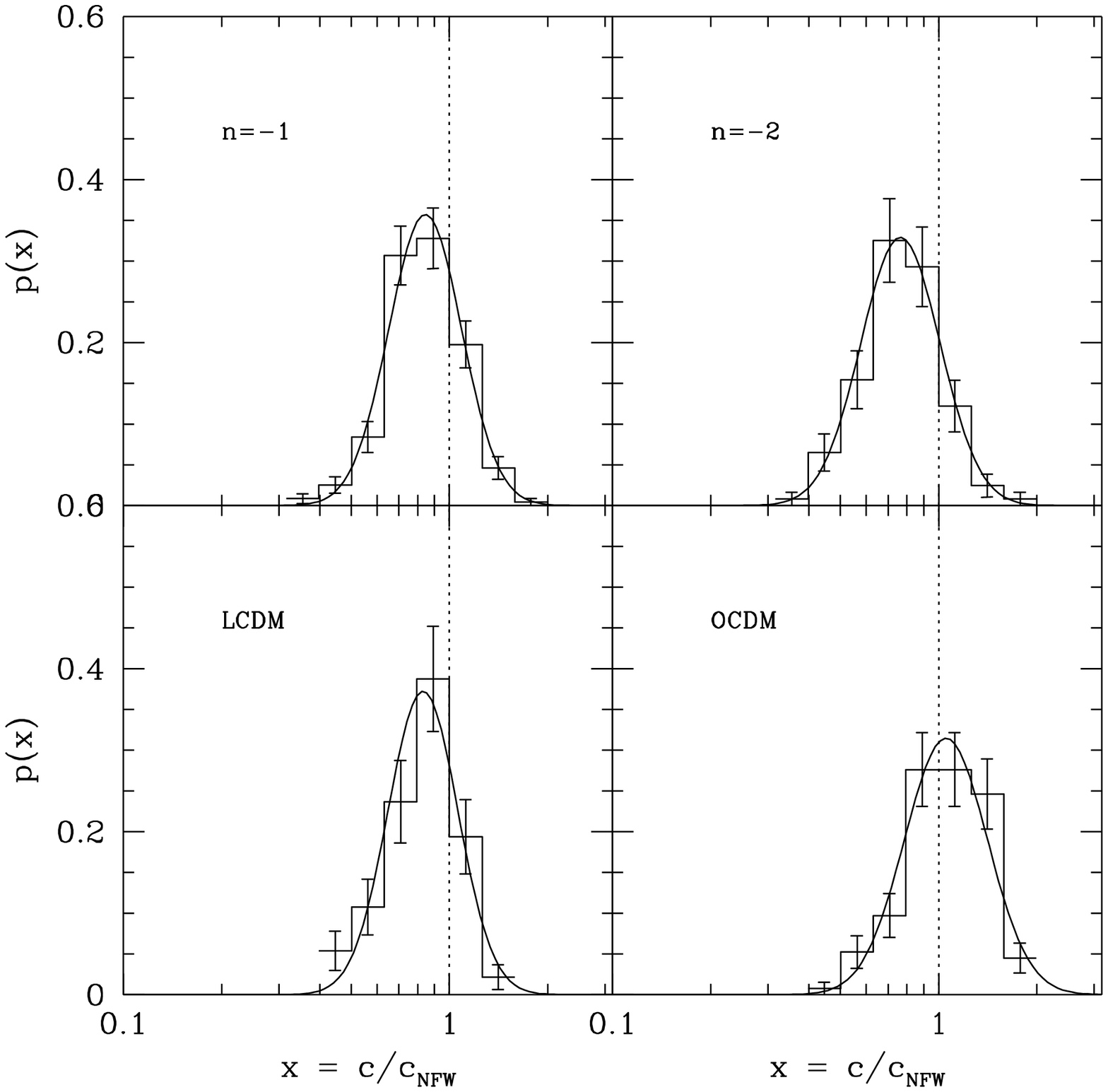}
\end{figure}

\begin{figure}
\epsscale{1.0} \plotone{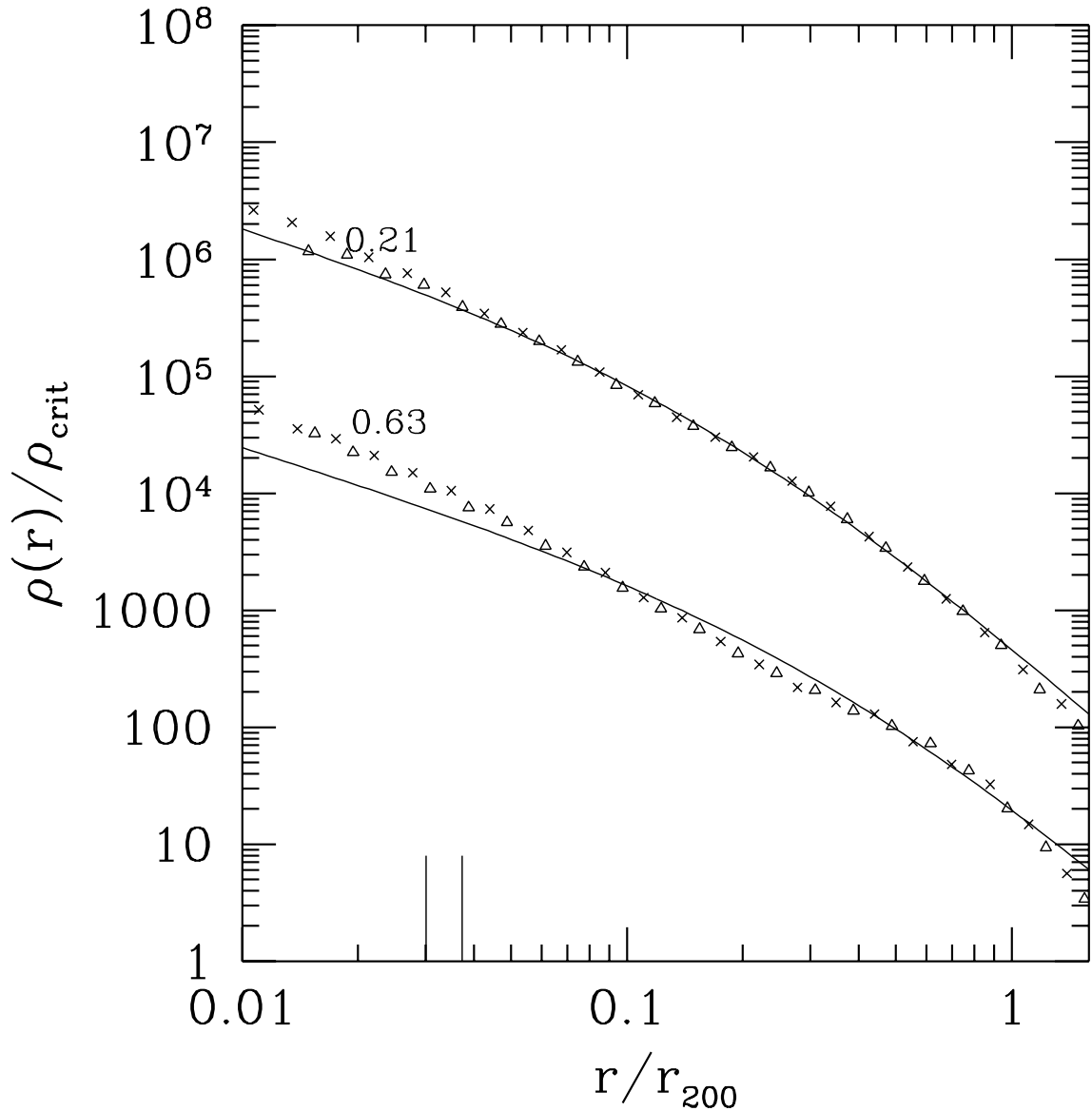}
\caption{The density profiles of the two halos in the LCDM model that
  are resimulated with a higher resolution, $\sim 10^6$ particles
  within the virial radius. The crosses show the result of these high
  resolution halo simulations, which should be compared with the
  profiles used for the present analysis (triangles; lower
  resolution). It is gratifying to see that the profiles of the
  different simulations agree very well for scales larger than the
  resolution limit (long ticks). The upper profiles have been vertically
  shifted up by a factor $10$.  }
\label{comp}\end{figure}

\newpage
%%%%%%%%%%%%%%%%%%%%%%%%%%%%%%%%%%%%%%%%%%%%%%%%%%%%%%%%%%%%%%%%%%%%%%%%%%
\begin{deluxetable}{cccccccccccccc}
  \tablecaption{ Simulations of $256^3$ particles} \tablewidth{0pt}
  \tablehead{ \colhead{Model} &\colhead{$\Omega_0$}
    &\colhead{$\lambda_0$} &\colhead{ $\Gamma$ or $n$
      \tablenotemark{a}} &\colhead{$\sigma_8$ or $M_\ast$
      \tablenotemark{b}} &\colhead{$L$\tablenotemark{c} }&\colhead{
      Num.\tablenotemark{d}}} \startdata SCDM & 1.0 & 0.0 & 0.5 & 0.85
  & 100 & 3\nl LCDM & 0.3 & 0.0 & 0.25 & 1.0 & 100& 3\nl OCDM & 0.3 &
  0.7 & 0.20 & 1.0 & 100&3\nl SF1& 1.0 & 0.0 &$-0.5$&18882&&3\nl SF2&
  1.0 & 0.0 &$-1.0$&14827&&3\nl SF3& 1.0 & 0.0 &$-1.5$&8832&&3\nl SF4&
  1.0 & 0.0 &$-2.0$&7834&&3\nl 
\enddata 
\tablenotetext{a} {The shape
    of the linear power spectrum. The CDM models are specified by
    $\Gamma$, and the scale-free models by $n$} 
\tablenotetext{b}{The
    amplitude of the linear power spectrum at the final epoch. The
    scale-free models are specified by $M_\ast$ and the CDM models by
    $\sigma_8$. $M_\ast$ is in units of the particle mass}
\tablenotetext{c}{In $\mpc$} 
\tablenotetext{d}{Number of independent
    realizations}
\end{deluxetable}
%%%%%%%%%%%%%%%%%%%%%%%%%%%%%%%%%%%%%%%%%%%%%%%%%%%%%%%%%%%%%%%%%%%%%%%%%%

\clearpage

\begin{deluxetable}{cccccccccccccc}
\tablecaption{Number of halos $N_h$ and the fitting results of $\bar c$ and
  $\sigma_c$ for a fixed range ${\rm dvi_{max}}$ \tablenotemark{a}}
\tablewidth{0pt}
\tablehead{
\colhead{Model}&\colhead{$0<{\rm dvi_{max}}<0.15$}&\colhead{$0.15<{\rm dvi_{max}}<0.30$}
&\colhead{$0.0<{\rm dvi_{max}}<0.30$}&\colhead{${\rm dvi_{max}}>0.30$}}
\startdata
$n=-0.5$& 70~~ 0.80~~ 0.21& 257~~ 0.84~~ 0.32 & 327~~ 0.84~~ 0.30&
352~~ 0.65~~ 0.50\nl
$n=-1$& 79~~ 0.93~~ 0.27& 238~~ 0.84~~ 0.26& 317~~ 0.86~~ 0.26&
246~~ 0.61~~ 0.44\nl
$n=-1.5$ & 84~~ 0.96~~ 0.18& 182~~ 0.79~~ 0.30& 266~~ 0.84~~ 0.28&
156~~ 0.59~~ 0.38\nl
$n=-2$ & 76~~ 0.91~~ 0.23& 123~~ 0.76~~ 0.28& 199~~ 0.81~~ 0.27& 96~~
0.55~~ 0.41\nl
SCDM & 57~~ 0.99~~ 0.18& 159~~ 0.90~~ 0.20& 216~~ 0.92~~ 0.20& 
119~~ 0.63~~ 0.28\nl
LCDM& 104~~ 0.95~~ 0.17& 93~~ 0.83~~ 0.25& 197~~ 0.90~~ 0.20& 40~~
0.55~~ 0.33\nl
OCDM& 101~~ 1.25~~ 0.17& 134~~ 1.05~~ 0.29& 235~~ 1.10~~ 0.23& 65~~
0.87~~ 0.39\nl
\enddata
\tablenotetext{a}{Three numbers, for example 70~~ 0.80~~ 0.21, in each 
  column are $N_h$, $\bar c$, and $\sigma_c$ respectively}
\end{deluxetable}

\end{document}